\documentstyle[graphicx,float]{article}

\begin{document}
\title{Absorption properties of identical atoms}
\date{}
\author{Pedro Sancho \\ Centro de L\'aseres Pulsados, CLPU \\ E-37008, Salamanca, Spain}
\maketitle
\begin{abstract}
Emission rates and other optical properties of multiparticle systems
in collective and entangled states differ from those in product
ones. We show the existence of similar effects in the absorption
probabilities for (anti)symmetrized states of two identical atoms.
The effects strongly depend on the overlapping between the atoms and
differ for bosons and fermions. We propose a viable experimental
verification of these ideas.
\end{abstract}
\vspace{7mm}


\section{Introduction}

The emission properties of multi-particle systems depend on the
quantum state of the system. The first example of this behavior was
presented by Dicke in his work on superradiance \cite{Dic,Fic,Har}:
a system in a collective state can radiate faster. The presence of
collective states does not only change the emission properties, but
also the absorption ones \cite{Rai}. Later, it has been both
theoretically and experimentally established the existence of
modifications of the emission properties in the case of initially
entangled states \cite{Fic,yo1,jap,yo2}. Other optical properties of
atoms, such as Raman scattering, can also depend on the entangled
character of the initial state \cite{Aga}. In a related context, it
has been demonstrated that the two-photon absorption rate can be
modified when the photons are entangled \cite{Fra}.

In this work we show that another type of multi-particle state, the
(anti) symmetrized state of identical particles, also leads to
modifications of the absorption properties. In particular, we derive
the absorption probabilities for systems of two identical particles,
which in some cases differ from those corresponding to product
states. These  differences are essentially determined by the
overlapping between the two atoms. Moreover, these modifications of
the absorption probabilities are not similar for bosons and
fermions.

We propose a scheme for the experimental verification of the effect
discussed in this work. A qualitative verification (involving a
large number of atoms instead of only two) can, in principle, be
obtained with minor modifications of existent techniques in the
fields of atom lasers and (anti)bunching verification. The
two-particle case could be experimentally studied with the same type
of technique, but it is more demanding.

\section{Two-particle states}

In this section we determine the initial and final states of the
atoms involved in the absorption process. We shall restrict our
considerations to two-atom systems. The two atoms interact with
light, for instance, a laser beam. We assume the intensity of the
laser not to be too high. This way, only one-photon absorptions are
relevant and we can neglect multi-photon processes.

The state of the atoms can be expressed as $|\psi _g>=|\psi >|g>$ or
$|\psi _e>=|\psi >|e>$, with $|\psi >$ denoting the center of mass
(CM) state, and $|g>$ and $|e>$ representing the ground and excited
states of the internal variables of the atom. We assume the laser to
be tuned to the frequency associated with that transition. With this
notation the initial state of the two-particle system is
\begin{eqnarray}
|\Psi _i>= N_i (|\phi _g>_1|\psi _g>_2 \pm  |\psi _g>_1|\phi _g>_2)=
\nonumber \\ N_i (|\phi >_1|\psi >_2 \pm  |\psi >_1|\phi >_2) |g
>_1|g >_2
\end{eqnarray}
with the upper and lower signs in the double sign expression
corresponding respectively to bosons and fermions. The normalization
factor is given by
\begin{equation}
N_i=\frac{1}{(2(1\pm |<\phi |\psi >|^2))^{1/2}}
\end{equation}
Let us now derive the final state after the absorption process. We
consider first the case where the particle in state $|\phi _g>$
absorbs one photon. There are two changes in the atom. On the one
hand, the internal state $|g>$ goes to the excited one $|e>$. On the
other hand, the state of the center of mass also changes as $|\phi >
\rightarrow |\tilde{\phi }>$. This process is a consequence of the
recoil of the atom. Similarly, for the other state we have the
evolution $|\psi _g > \rightarrow |\tilde{\psi }_e>$ after the
photon absorption. Thus, the final state must contain $\tilde{\phi
}_e$ and $\tilde{\psi }_e$ as the excited states, and $\phi _g$ and
$\psi _g$ as the non-excited ones.

When the two final CM states are different, $\tilde{\phi }_e$ and
$\tilde{\psi }_e$ are distinguishable alternatives to describe the
two-particle system with one of the atoms in an excited state. After
the absorption the system is not in a superposition of the two
alternatives but in a mixture of the states representing them:
$|\Psi _f(\tilde{\phi })>$ and $|\Psi _f (\tilde{\psi })>$, where
the symbol between parentheses indicates the final CM state of the
excited atom. Their explicit expressions are
\begin{equation}
|\Psi _f(\tilde{\phi })>= \frac{1}{\sqrt{2}} (|\tilde{\phi }_e>_1|\psi _g>_2 \pm  |\psi _g>_1 |\tilde{\phi }_ e >_2)
\end{equation}
and
\begin{equation}
|\Psi _f(\tilde{\psi })>= \frac{1}{\sqrt{2}} (|\tilde{\psi }_e>_1|\phi _g>_2 \pm  |\phi _g>_1 |\tilde{\psi }_ e >_2)
\end{equation}

There is a particular case that must be considered separately, that
when the final states are equal, $|\tilde{\psi }> = |\tilde{\phi }>$
(because of the Pauli exclusion principle this case only refers to
bosons). This situation  clearly also corresponds to equal initial
states, $|\psi > = |\phi >$. We cannot know if the absorption
process has taken place via the particle labeled as $1$ or that as
$2$. These two alternatives are indistinguishable and the final
state of the system must be a superposition of the states
representing them:
\begin{equation}
|\Psi _f (eq)>= \frac{1}{\sqrt{2}} (|\tilde{\phi }_e>_1|\phi _g>_2 +  |\phi _g>_1 |\tilde{\phi }_ e >_2)
\label{eq:cin}
\end{equation}
This form automatically fulfills the symmetrization condition. The
fundamental difference with the case of non equal CM states is that
now we have a pure state describing the two-particle system whereas
previously we had a mixture.

\section{Absorption probabilities}

After deriving the initial and final states of the two-atom system
we can easily obtain the matrix elements and probabilities for the
absorption. The transition matrix elements for this process can be
evaluated as
\begin{equation}
{\cal M}_{\cal F}=<\Psi _{\cal F}|\hat{U}_{12}|\Psi _i>
\end{equation}
where $|\Psi _{\cal F}>$ denotes any of the possible final states
and $\hat{U}_{12}$ is the evolution operator of the two-particle
system. Note that for the matter of simplicity we have not included
in the above expression the variables associated with the
electromagnetic field. In the initial state we should add
$|n>_{EM}$, and in the final one $|n-1>_{EM}$. In the usual dipole
and rotating-wave approximations \cite{Lou}, and to first order of
perturbation theory the complete matrix element would read
proportional to $<\Psi _{\cal F}|\hat{U}_{12}(\hat{\bf D}_1,
\hat{\bf D}_2 )|\Psi _i> <n-1|\hat{\bf E}|n>_{EM}$ with $\hat{\bf
E}$ the electric field operator, and $\hat{\bf D}_1$ and $\hat{\bf
D}_2$ the dipole operators.

The evolution operator can be expressed as
\begin{equation}
\hat{U}_{12}=\hat{U}_1 \otimes \hat{U}_2
\end{equation}
with $\hat{U}_i$, $i=1,2$, the evolution operators of the particles.
In order to simplify the problem and to capture the main physical
ideas we neglect the atom-atom interactions. Then we can take
$\hat{U}_i$ as the evolution operator of the particle only in
interaction with the radiation field. Moreover, as the two particles
are identical the two evolution operators are equal
$\hat{U}_1=\hat{U}_2= \hat{U}$, and we can write
$\hat{U}_{12}=\hat{U} \otimes \hat{U}$.

Note that we do not explicitly include in $\hat{U}_{12}$ the initial
and final times of the evolution ($\hat{U}_{12}(t_f,t_i)$). We take
$t_f-t_i$ as the  duration of the laser pulse in each repetition of
the experiment, which we assume to be short enough to avoid multiple
absorption processes (such as absorption-spontaneous
emission-absorption or two absorptions processes) that would
complicate the description of the system.

Let us compare the probabilities of having the particles in the
final states $\tilde{\phi }_e$ and $\psi _g$ after the absorption
process for two-atom systems in (anti)symmetrized and product
states. The transition matrix element for the (anti)symmetrized
state is ${\cal M}_{\tilde{\phi}}=<\Psi _f(\tilde{\phi
})|\hat{U}_{12}|\Psi _i>$, which can be expressed as
\begin{equation}
{\cal M}_{\tilde{\phi }}= \frac{<\tilde{\phi}_e|\hat{U}| \phi _g> <\psi _g|\hat{U}| \psi _g> \pm <\tilde{\phi}_e|\hat{U}|\psi _g><\psi _g|\hat{U}|\phi _g>}{(1\pm |<\phi |\psi >|^2)^{1/2}}
\label{eq:uno}
\end{equation}
This probability amplitude contains two contributions, the direct one associated with the evolution
\begin{equation}
\phi _g \rightarrow \tilde{\phi }_e \; ; \; \psi _g \rightarrow \psi
_g \label{eq:ee1}
\end{equation}
and the crossed one representing the alternative evolution
\begin{equation}
\psi _g \rightarrow \tilde{\phi }_e \; ; \; \phi _g \rightarrow \psi
_g \label{eq:eed}
\end{equation}
All the matrix elements in the numerator of Eq. (\ref{eq:uno})
correspond to the single-particle probability amplitudes for the
evolutions in expressions (\ref{eq:ee1}) and (\ref{eq:eed}). Then when the
probability is evaluated, we have
\begin{eqnarray}
(1\pm |<\phi |\psi >|^2) P_{two}(\tilde{\phi })= P_{sin}(\phi _g \rightarrow \tilde{\phi }_e) P_{sin}(\psi _g \rightarrow \psi _g) + \nonumber \\
P_{sin}(\psi _g \rightarrow \tilde{\phi }_e) P_{sin}(\phi _g
\rightarrow \psi _g) \pm \nonumber \\
2Re({\cal M}_{sin}^*(\phi _g \rightarrow \tilde{\phi }_e) {\cal
M}_{sin}^*(\psi _g \rightarrow \psi _g) {\cal M}_{sin}(\psi _g
\rightarrow \tilde{\phi }_e) {\cal M}_{sin}(\phi _g \rightarrow \psi
_g))
\end{eqnarray}
where $ P_{two}(\tilde{\phi })= |{\cal M}_{\tilde{\phi }} |^2$ is
the probability of the two-particle system to be in the final state
$ \Psi_f(\tilde{\phi })$ when starting in $\Psi _i$, $P_{sin}(\phi _g \rightarrow \tilde{\phi }_e) $ represents the probability of a single atom evolving from $\phi _g$ to $\tilde{\phi }_e$,... The last term in the
r. h. s. represents the interference effects between the two
alternatives. This is a manifestation of the exchange effects.

There is a particular scenario where we can have a more clear
picture. It corresponds to the case when the second term in the
numerator of Eq. (\ref{eq:uno}) can be neglected in comparison with
the first one. This situation corresponds, for instance, to
evolutions such that the evolved $\phi$ without absorption is very
different from $\psi$.

In this scenario the matrix element simplifies to
\begin{equation}
{\cal M}_{\tilde{\phi}} \approx \frac{<\tilde{\phi}_e|\hat{U}| \phi _g> <\psi _g|\hat{U}|\psi _g>}{(1\pm |<\phi |\psi >|^2)^{1/2}}
\end{equation}
We want to compare this probability amplitude with that of a pair of
identical atoms in factorized states. In this case the initial
state is $|\Psi _i^{fac}>= |\psi _g>_1 |\phi _g>_2$, and the final
one $|\Psi _f^{fac}(\tilde{\phi })>= |\psi _g>_1 |\tilde{\phi }
_e>_2$. We have the transition probability amplitude ${\cal
M}_{\tilde{\phi}}^{fac} =<\Psi _f^{fac}(\tilde{\phi
})|\hat{U}_{12}|\Psi _i^{fac}>$, which leads to the probability
\begin{equation}
P_{two}^{fac}(\tilde{\phi })= P_{sin}(\phi _g \rightarrow \tilde{\phi }_e)P_{sin}(\psi _g \rightarrow \psi _g)
\end{equation}
with $ P_{two}^{fac}(\tilde{\phi })$ denoting the probability of the
two-particle system to be in the final state $
\Psi_f^{fac}(\tilde{\phi })$ when initially was in $ \Psi_i^{fac}$.

The ratio of the two probabilities is
\begin{equation}
\frac{P_{two}(\tilde{\phi })}{P_{two}^{fac}(\tilde{\phi })}  \approx \frac{1}{1\pm |<\phi |\psi >|^2}
\end{equation}
The ratio strongly depends on the initial overlapping of the two
atoms (such as measured by $|<\phi |\psi >|^2$). We observe an
opposite behaviour for bosons and fermions. In the first case, the
absorption probability in the symmetrized state is smaller than in
the product one. In contrast, for fermions, the ratio is larger than
unity.

In the limit of negligible overlapping these effects tend to
disappear, $P_{two}(\tilde{\phi }) \approx P_{two}^{fac}(\tilde{\phi })$.

We have only considered the problem of absorption of a photon with
final states $\tilde{\phi }$ and $\psi$. The case of absorption with
final states $\tilde{\psi }$ and $\phi$ is similar. Finally, the
case in which the absorption can give rise to any of the pairs of
final states, also easily follows taking into account that the two
final states are distinguishable and the probabilities add.

When the second term in the numerator of Eq. (\ref{eq:uno}) is
comparable to the first one, one must deal with the complete
equation. This situation occurs, for instance, for large
overlappings. We shall only consider the limiting case of very
similar initial CM wavefunctions, $\psi \approx \phi$. In this case
we have $|<\phi |\psi >|^2 \approx 1$ and ${\cal M}_{\tilde{\phi}}
\approx (1\pm 1) <\tilde{\phi}_e|\hat{U}| \phi _g> <\psi
_e|\hat{U}|\psi _g>/(1 \pm 1)^{1/2}$. For fermions this expression
is undefined, reflecting the Pauli principle. On the other hand, for
bosons gives $P_{two}(eq )=2 P_{sin}(\phi _g \rightarrow
\tilde{\phi}_e) P_{sin}(\phi _g \rightarrow \phi _g)$, expressing $
P_{two}(eq )$ the probability of one absorption in the two-particle
system when the two wavefunctions of the symmetrized initial state
are equal. For a factorized two-particle state we have that the
absorption can happen for $|\phi _g>_1$ and $|\phi _g>_2$. Then we
have $P_{two}^{fac}(eq )=2 P_{sin}(\phi _g \rightarrow
\tilde{\phi}_e) P_{sin}(\phi _g \rightarrow \phi _g)$ and
$P_{two}(eq)=P_{two}^{fac}(eq)$, with $ P_{two}^{fac}(eq )$ the
probability of one absorption in the two-particle system when the
two initial wave functions are equal and the initial and final
states are factorized ones. In the limit of equal initial CM states
there are not differences between symmetrized and factorized states.
Note that the last result can also be derived starting directly from
the state $|\Psi _f(eq)>$ introduced in Eq. (\ref{eq:cin}), which
gives ${\cal M}_{eq}=\sqrt{2} <\tilde{\phi}_e|\hat{U}|\phi _g> <\phi
_g|\hat{U}|\phi _g>$

\section{A proposal for an experimental test}

We discuss in this section if the effects considered above could be
tested experimentally. We propose a scheme where it is possible, in
principle, to carry out an experimental verification of the problem.
Our scheme closely follows some ideas of the arrangements used in
atom laser experiments and the verification of (anti)bunching
effects (see, for instance, \cite{Ess} for the first field and
\cite{Asp} for the second). We shall discuss separately two versions
of the scheme. The first one only could provide a qualitative
verification, but almost does not imply modifications of the part of
the arrangement \cite{Ess} that we use. The second one is much more
demanding from the technical point of view, but could test the
two-atom case.

\subsection{The qualitative verification}

As in \cite{Ess}, we have a large number of atoms ($^{87}Rb$) in a
magnetic trap. The atoms are in an hyperfine ground state
($|F=1,m_F=-1>$). A beam of microwave light tuned to the transition
$|F=1,m_F=-1> \rightarrow |F=2,m_F=0>$ is focused on the trap. The
absorption of the microwave photons spin-flips atoms into the final
state, which do not longer experience the trapping potential and can
escape. The released particles fall by the effect of gravity and,
downwards they meet atom detectors. These detectors can be of the
type considered in \cite{Asp} (their relevant properties for our
experiment will be discussed later). Counting the number of atoms
arriving to the detectors we can infer the number of absorptions
that have taken place in the trap. This way we have a simple method
to measure absorptions. With a very large attenuation of the
intensity of the light beam we can expect very few absorptions for
each microwave pulse.

A second fundamental element for the scheme is the control of the
degree of overlapping in the trap. As signaled before, the effects
discussed in this paper depend on the overlapping between the
identical particles. As it is well-known in Bose-Einstein
condensation or ultracold gases theory the degree of overlapping is
related to the thermal wavelength, $\lambda _T =h/(2\pi mkT) ^{1/2}$
with $T$ the temperature and $k$ Boltzmann's constant. Lowering the
temperature of the ensemble of atoms we can make $\lambda _T$,
directly related to de Broglie's wavelength, comparable to the mean
distance between atoms. In this case we hope the proposed effect to
manifest. In contrast, for higher temperatures, the overlapping will
be in general small and the particles will behave as in product
states.

Thus, comparing the number of absorptions in the trap for different
temperatures we should observe the modifications of the absorption
probabilities as a function of temperature. This verification would
correspond to a multi-atom system, not to a two-atom one as
discussed in this paper. In consequence, the simple mathematical
description given here should be completed taking into account
multi-particle symmetrization, or the possibility of collective
effects such as those discussed in \cite{Rai}. Depending on the
results of that extended analysis, the proposed scheme could be a
verification of the dependence of the absorption probabilities on
the overlapping, as this depends on the temperature.

\subsection{The quantitative verification}

An extension of the above scheme could deal, in principle, with the
two-atom system discussed above. We present separately the different steps of the scheme (see Fig.1).
\begin{figure}[H]
\center
\includegraphics[width=7cm,height=7cm]{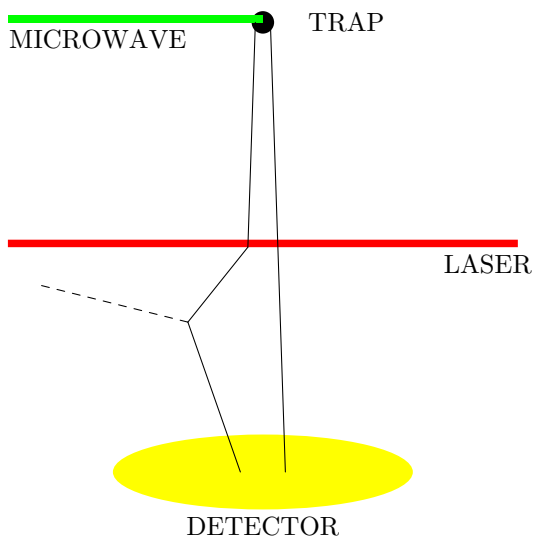}
\caption{The continuous lines leaving the trap represent two
released atoms. One of them absorbs in the interaction with the
laser, emitting spontaneously in the subsequent evolution a photon
(discontinuous line). The number of emissions can be counted
collecting the photons with lenses and driving them to light
detectors.}
\end{figure}

{\bf Preparation of the initial state.} As before, a microwave beam
releases particles from the trap. The intensity of the beam is
attenuated in such a way that only very few atoms (ideally only one)
escape with each pulse. In each repetition of the experiment two
pulses with a very short tunable delay between them are directed
towards the trap. Modulating the temporal delay, we can control the
overlapping.

With a postselection process that we shall discuss below we can
restrict our considerations to the case of having two released atoms
in each run of the experiment.

{\bf Interaction with the laser.} Later, the atoms interact with
another laser (see Fig. 1). The frequency of this laser is tuned to
a transition (not the hyperfine one considered before, which now
only has the role of releasing atoms from the trap) of the atom. The
intensity of the laser is low enough to only have large
probabilities of one-photon absorption.

Note that being the two atoms released at slightly different times,
their states at the time of interaction with the laser will be
slightly different, depending the overlapping of the delay between
the microwave beams.

{\bf Spontaneous emission.} After the interaction, the atoms
continue their downwards propagation. We choose the distance between
the interaction region and the detectors long enough for most of the
excited atoms return to the ground state by spontaneous emission
before reaching the atom detectors. This condition can be easily
evaluated using the lifetime of the excited state and the mean
velocity of the atoms. By placing photon detectors we can determine
the number of spontaneous emission events. In the final step of our
scheme the atoms reach the atom detectors. At this stage we must
carry out the postselection process to single out the cases with two
atoms escaping from the trap. As signaled before we use atom
detectors as in \cite{Asp}. They allow for position- and
time-resolved detection events. The second aspect is fundamental
because, introducing a temporal window, we can determine if the
events have been generated by the two microwave pulses of a single
run. The number of spontaneous emissions gives us the number of
absorptions in the laser-released atoms interaction. Next, we must
compare these numbers for different overlappings between the
released atoms. As discussed before, the overlapping can be
controlled via the temporal delay between the two microwave pulses.
A comparison of the number of absorptions for different delays could
test the effects predicted in this paper. Note, as discussed at the
end of Sect. 3, that in the case of very similar initial wave
functions (very small delays) the changes of the absorption rate are
very small and probably undetectable from the experimental point of
view. It is for this reason that our proposal to observe the effect
focus on the dependence of the absorption rate on the overlapping
degree, rather than on the determination of the rate for particular
values of the overlapping.

\section{Discussion}

We have shown how the absorption properties of a pair of identical
atoms are related to the (anti)symmetric properties of its state.
This is another example of the dependence of optical properties of
quantum systems on their quantum states.

The process resembles in some aspects the superradiance phenomenon,
but we must also emphasize the differences. In superradiance the
collective state leads to strong correlations between the atomic
dipoles, which are in the basis of the emission enhancement. The
collective states originally introduced by Dicke had a high degree
of symmetry: the atom-radiation description was invariant under atom
permutations. This closely resembles our approach, where the
(anti)symmetry of the states of the identical atoms generates the
modifications of the absorption properties. However, in the case of
superradiance, it was later realized that a more realistic
description of the process must introduce interactions that break to
some extent the permutational invariance \cite{Har}. Another
difference is that for collective states the properties are
independent of the identity (bosonic or fermionic) of the atoms,
being only a function of the separation between atoms,....

We also must compare our approach with that associated with
entangled states \cite{Fic,yo1,jap,yo2}. In this case the two
mathematical formalisms are similar due to the resemblances of the
states describing both systems. As a matter of fact, the
(anti)symmetric states are formally identical to entangled states.
However, the physical mechanisms underlying both types of
non-factorizable states are essentially different. In our case we
need proximity between both particles (the overlapping cannot be
negligible), whereas the effects associated with entanglement can
manifest at large distances. In addition, we can have entangled
states giving super- or sub-radiance \cite{Fic} independently of the
atoms identity.

We have proposed an arrangement that could test the existence of the
effects discussed in this paper. The simpler version of the
arrangement is a simple modification of techniques used in other
experiments. It has the inconvenience of only giving a qualitative
verification of the effects. The more elaborated version, where the
dependence on the initial overlapping could be analyzed
experimentally, seems to be more difficult to implement.

{\bf Acknowledgments} I acknowledge support from Spanish Ministerio
de Ciencia e Innovaci\'on through the research project
FIS2009-09522.

\end{document}